\documentclass[amssymb,nobibnotes,superscriptaddress,longbibliography,notitlepage,twocolumn]{revtex4-2}
\usepackage[english]{babel}
\usepackage[utf8]{inputenc}
\usepackage[colorinlistoftodos, color=green!40, prependcaption]{todonotes}
\usepackage{braket}
\usepackage{bm}
\usepackage{graphicx}
\usepackage{multirow}
\usepackage{bbold}
\usepackage{amsthm}
\usepackage{mathtools}
\usepackage{physics}
\usepackage{xcolor}
\usepackage{graphicx}
\usepackage[left=23mm,right=13mm,top=35mm,columnsep=15pt]{geometry} 
\usepackage{adjustbox}
\usepackage{placeins}
\usepackage[T1]{fontenc}
\usepackage{lipsum}
\usepackage{csquotes}

\usepackage[pdftex, pdftitle={Article}, pdfauthor={Author}]{hyperref} % For hyperlinks in the PDF

\begin{document}
\title{Quantum tangent kernel}

\author{Norihito Shirai}
\email{shirai@qc.ee.es.osaka-u.ac.jp}
\affiliation{Graduate School of Engineering Science, Osaka University, 1-3 Machikaneyama, Toyonaka, Osaka 560-8531, Japan.}

\author{Kenji Kubo}
\email{kenjikun@mercari.com}
\affiliation{Graduate School of Engineering Science, Osaka University, 1-3 Machikaneyama, Toyonaka, Osaka 560-8531, Japan.}
\affiliation{R4D, Mercari Inc., Roppongi Hills Mori Tower 18F, 6-10-1, Roppongi, Minato-ku, Tokyo 106-6118, Japan}

\author{Kosuke Mitarai}
\email{mitarai@qc.ee.es.osaka-u.ac.jp}
\affiliation{Graduate School of Engineering Science, Osaka University, 1-3 Machikaneyama, Toyonaka, Osaka 560-8531, Japan.}
\affiliation{Center for Quantum Information and Quantum Biology, Osaka University, Japan.}
\affiliation{JST, PRESTO, 4-1-8 Honcho, Kawaguchi, Saitama 332-0012, Japan.}

\author{Keisuke Fujii}
\affiliation{Graduate School of Engineering Science, Osaka University, 1-3 Machikaneyama, Toyonaka, Osaka 560-8531, Japan.}
\affiliation{Center for Quantum Information and Quantum Biology, Osaka University, Japan.}
\affiliation{RIKEN Center for Quantum Computing, Wako Saitama 351-0198, Japan}

\date{\today}

\begin{abstract}
Quantum kernel method is one of the key approaches to quantum machine learning, which has the advantages that it does not require optimization and has theoretical simplicity.
%Quantum machine learning models are thought to be interpretable as quantum kernel method.
By virtue of these properties, several experimental demonstrations and discussions of the potential advantages have been developed so far.
However, as is the case in classical machine learning,
not all quantum machine learning models could be regarded as kernel methods.
In this work, we explore a quantum machine learning model with a deep parameterized quantum circuit and aim to go beyond the conventional quantum kernel method. 
In this case, the representation power and performance are expected to be enhanced, while the training process might be a bottleneck because of the barren plateaus issue.
%The kernel calculated by using the deep quantum circuit outperforms the conventional quantum kernel method for the classification task on the ansatz-generated dataset.
However, we find that parameters of a deep enough quantum circuit do not move much from its initial values during training, allowing first-order expansion with respect to the parameters.
This behavior is similar to the neural tangent kernel in the classical literatures,
and such a deep variational quantum machine learning can be described by another emergent kernel, {\it  quantum tangent kernel}.
Numerical simulations show that the proposed quantum tangent kernel outperforms the conventional quantum kernel method for an ansatz-generated dataset.
%The kernel calculated by using the deep quantum circuit outperforms the conventional quantum kernel method for the classification task on the ansatz-generated dataset.
%We call the kernel calculated by using this quantum circuit quantum tangent kernel (QTK).
This work provides a new direction beyond the conventional quantum kernel method and explores potential power of quantum machine learning with deep parameterized quantum circuits.
\end{abstract}

%\keywords{first keyword, second keyword, third keyword}

\maketitle

\section{Introduction} \label{sec:introduction}
% Variational quantum algorithms (VQAs) \cite{Cerezo2021} has emerged as a possible application of noisy intermediate scale quantum (NISQ) devices \cite{Preskill2018quantumcomputingin}.
% VQAs are the family of quantum algorithm which utilize parameterized quantum circuit and optimize the parameters for performing certain tasks such as producing ground states of quantum system.

Applying quantum computers to machine learning purposes is an emerging area of research.
In particular, motivated by the recent advance of hardware technology \cite{Arute2019}, techniques to apply so-called noisy intermediate quantum (NISQ) devices \cite{Preskill2018quantumcomputingin} are rapidly developed \cite{mitarai2018quantum,havlivcek2019supervised,Schuld2019feature,Benedetti2019pqc,Cerezo2021}.
One direction that is frequently explored is a variational method which use parameterized quantum circuits to construct a model $y(\bm{x}, \bm{\theta})$ which outputs a prediction when fed with an input data $\bm{x}$.
More specifically, using a parameterized quantum circuit $U(\bm{x},\bm{\theta})$ with trainable parameters $\bm{\theta}$, we construct a model $y(\bm{x},\bm{\theta})$ for an input $\bm{x}$ by the expectation value of an observable $O$: $y(\bm{x}, \bm{\theta})=\bra{0}U^\dagger(\bm{x},\bm{\theta}) O U(\bm{x},\bm{\theta})\ket{0}$.
Since there are quantum circuits that are hard to simulate classically and can be implemented on present hardware \cite{Arute2019}, using such types of circuit for $U(\bm{x},\bm{\theta})$ we might be able to construct a machine learning model that exceeds the capability of classical computers.

Another promising direction is the so-called quantum kernel method \cite{Schuld2019feature,havlivcek2019supervised,schuld2021supervised}.
In this approach, we use quantum computers solely for preparing a classically intractable feature $\ket{\phi(\bm{x})}$ and taking the inner products of the feature vectors.
The training is performed on a classical computer using the values of inner product between each pair of training dataset.
It is advantageous in that we do not have to optimize the quantum circuit, which is often difficult in the aforementioned variational methods.
Moreover, it can be shown that the quantum kernel method outperforms the variational ones in a certain sense \cite{schuld2021supervised}; if $U(\bm{x},\bm{\theta})$ takes the form of $U(\bm{x},\bm{\theta}) = V(\bm{\theta})U_{\phi}(\bm{x})$, then the quantum kernel method with feature vector $\ket{\phi(\bm{x})}=U_{\phi}(\bm{x})\ket{0}$ is guaranteed to achieve better prediction accuracy for training dataset.
Its simple framework greatly advanced the construction of quantum machine learning theory, providing insights on potential advantages~\cite{huang2021power,liu2021rigorous}.
Also, owing to its experimental easiness, there have been several experimental demonstrations of the method \cite{Kusumoto2021,Bartkiewicz2020,glick2021covariant}.
However, as is well known in the machine learning field, not all machine learning models can be described by kernel methods, and there must be approaches that go beyond them, such as deep learning.

In this work, we explore how to construct a quantum machine learning model that goes beyond the conventional quantum kernel method.
To this end, we investigate the performance of a model constructed by deep quantum circuit in the form of $U(\bm{x},\bm{\theta}) = \prod_i V_i(\bm{\theta}_i)U_{\phi, i}(\bm{x})$.
The circuit is now not in the form of $U(\bm{x},\bm{\theta}) = V(\bm{\theta})U_{\phi}(\bm{x})$, and hence this model cannot be translated to a kernel framework.
Unfortunately, as parameterized quantum circuits become deeper and go beyond the conventional quantum kernel method, it becomes more difficult to train the parameters as was the case with deep neural network in the classical literature.
While this issue would be resolved by finding a good ansatz and/or initial parameters \cite{Cong2019,PhysRevX.11.041011,Grant2019initialization}, we consider an alternative approach, i.e., overparameterization \cite{Kim2021,larocca2021theory,Wierichs2020}.

More precisely, we find that, when the circuit is deep enough, each of the parameters in $\bm{\theta}$ does not move much and stays close to the initial random guess.
This behavior is similar to the classical deep neural network \cite{DBLP:journals/corr/abs-1806-07572},
where the amount of change in the parameters is small, and the network is well described by a linear model with the tangent on the parameters as the basis function.
Thus optimized parameters on overparameterized networks can be found by another kernel, so-called neural tangent kernel.
The observation motivates us to propose quantum tangent kernel (QTK), which is a quantum analog of neural tangent kernel \cite{DBLP:journals/corr/abs-1806-07572}.

Specifically, we define QTK by the kernel associated with a feature map $\bm{x}\to\nabla_{\bm{\theta}}y(\bm{x},\bm{\theta})$ for 
a deep parameterized quantum circuit.
The QTK can be calculated efficiently on a quantum computer
by using analytical differentiation of parameterized quantum circuits, so-called parameter-shift rule~\cite{mitarai2018quantum, schuld2019evaluating}. 
Then, QTK combined with the standard kernel methods such as support vector machine (SVM) allows us to learn and infer without any explicit optimization of the parameterized quantum circuit.
We compare the performance of QTK and the conventional quantum kernel method for an ``ansatz-generated'' dataset generated by a deep parameterized quantum circuit, while the parameters are randomly chosen apart from those in the QTK.
As a result, we find that QTK outperforms the conventional quantum kernel method. This and the fact that tangent kernel is justified only in the overparameterized limit imply that deep parameterized quantum circuits have a great potential to go beyond the conventional quantum kernel model.
%We numerically show that the QTK can classify datasets that are hard to classify with an exisiting quantum kernel method. 
This work opens up a new direction beyond the conventional quantum kernel method for deeper quantum machine learning.

\section{Background} \label{sec:kernel}

\subsection{Kernel methods}
In this subsection, we explain the kernel method and support vector machine (SVM), which is one of the kernel-based classification methods.
Throughout this paper, we denote the training dataset by $\mathcal{D} = \{ (\bm{x}_i, y_i) \}_{i=1}^N $ where $\bm{x}_i$ is input data and $y_i$ is corresponding teacher data. 

The kernel method is a technique for mapping features to a higher dimensional feature space in order to introduce non-linearity to the model. 
%For example, in a classification problem, if the data cannot be linearly separated, non-linear mapping the data to a higher dimensional feature space may make linear separation possible. 
% The data that cannot be linearly separated can be linearly separated by applying a nonlinear transformation to a higher dimensional space.
It employs a non-linear map $\phi$ from the original space to the higher dimensional feature space:
\begin{align}
    \phi : \chi & \to \mathcal{H} \, , \\
    \bm{x}_i & \to \phi(\bm{x}_i) \, , \nonumber
\end{align}
where $\chi$ is an original space of the data and $\mathcal{H}$ is a higher dimensional feature space.
%However, as the dimensionality of $\mathcal{H}$ increases, calculating $\phi$ explicitly becomes computationally expensive.
% The performance highly depends on the choice of $\phi$ and finding the good $\phi$ itself becomes computationally expensive because of the curse of dimensionality.
In the kernel method, we only use inner products of $\phi$ with respect to different input data.
For two data $\bm{x}_i$ and $\bm{x}_j$, we define
\begin{align}
    K(\bm{x}_i, \bm{x}_j) = \langle \phi(\bm{x}_i) , \phi(\bm{x}_j) \rangle \, ,
\end{align}
where $\langle \cdot , \cdot \rangle$ denotes the inner product on $\mathcal{H}$.
The kernel method has the advantage that it does not require direct computation of the vectors in a high-dimensional feature space as long as their inner product can be calculated efficeintly.
% If only the inner product of vectors in a high-dimensional feature space appears in the objective function, then only the inner product needs to be calculated without directly computing the mapping to the feature space.
This inner product represents the similarity between the features and is called the kernel function.

As an example of the kernel method, we consider SVM, which is one of the kernel-based classification methods.
% The optimization problem of SVM can be formulated as dual quadratic problem.
SVM with kernel $K(\bm{x}_i, \bm{x}_j)$ is trained by maximizing the following objective function
\begin{align}
    L_D(\alpha) =  - \frac{1}{2} \sum_{i, j = 1}^N \alpha_i \alpha_j y_i y_j K(\bm{x}_i, \bm{x}_j) + \sum_{i = 1}^N \alpha_i \, ,
\end{align}
subject to $\sum_{i=0}^N \alpha_i y_i = 0$ and $\alpha_i \geq 0$.
$L_D$ depends only on the kernel function, and hence we do not need to explicitly calculate the feature map $\phi$.
The solution $\alpha^*_i$ to the above optimization problem is used for building the prediction model as follows:
\begin{align}
    y(\bm{x}) = {\rm sign} \left ( \sum_{i = 1}^N y_i \alpha^*_i K(\bm{x}_i, \bm{x}) + b \right ) \, ,
\end{align}
where the bias $b$ is calculated as,
\begin{align}
    b = y_j - \sum_{i=1}^N \alpha^*_i y_i K(\bm{x}_i, \bm{x}_j) \, .
\end{align}
Although the formula of bias $b$ holds for any $j$, practically, the average for all $j$ is taken.

% In this paper, we propose quantum tangent kernel (QTK) and evaluate the classification performance of SVM using it in section \ref{sec:numerical_experiment}.

\subsection{Neural Tangent Kernel} \label{subsec:NTK}
Let us denote by $f(\bm{x}_i, \bm{\theta})$ the output of a neural network where $\bm{\theta}$ is parameters in the network, and $\bm{x}_i$ is the input data. 
In the large width overparameterized neural network, the values of the parameters change only slightly from the initial values during training even if the initial values of parameters are set randomly  \cite{DBLP:journals/corr/abs-1806-07572}.
Therefore, the output of such a neural network can be well approximated by the first-order expansion with respect to the parameters around the initial values:
\begin{align}\label{eq:neural-network-linear-approx}
    f(\bm{x}_i, \bm{\theta}) \simeq f(\bm{x}_i, \bm{\theta}_0) + \nabla_{\theta} f(\bm{x}_i, \bm{\theta}_0)^T (\bm{\theta} - \bm{\theta}_0) \, ,
\end{align}
where $\bm{\theta}_0$ is the initial values of parameters of a neural network.
This approximation allows us to interpret the neural network as a linear model with a feature map
\begin{align}
    \phi(\bm{x}) = \nabla_{\theta} f(\bm{x}, \bm{\theta}_0) \, .
\end{align}
Using this feature map, we define the following kernel called neural tangent kernel (NTK) \cite{DBLP:journals/corr/abs-1806-07572, lee2019wide}.
%By using the least square error and the update formula of the gradient descent with an infinitesimal learning rate Eq.(\ref{eq:diff-gradient-descent}), the training dynamics of a large width neural network is described by the following  ordinary differential equation.
\begin{align}
%    \frac{d f(\vector{x}_i, \vector{\theta}(t))}{d t}
%    &= - \sum_{j=1}^{N} K_{\rm ntk}(\vector{x}_i, \vector{x}_j) \left ( f(\vector{x}_j, \vector{\theta}(t)) - y_i \right ) \, , \\
     K_{\rm ntk}(\bm{x}_i, \bm{x}_j) &= \nabla_{\theta} f(\bm{x}_i, \bm{\theta}_0)^T \nabla_{\theta} f(\bm{x}_j, \bm{\theta}_0) \, .
\end{align}
By its construction, NTK-based linear models are expected to be equivalent to the large-width neural network as long as the assumption Eq. (\ref{eq:neural-network-linear-approx}) is valid.
% Even if the parameters are trained explicitly  using such as gradient descent, in the large width limit, $K_{\rm ntk}(\vector{x}_i, \vector{x}_j)$ remains constant during training.
% It is called the neural tangent kernel (NTK) 

\subsection{Quantum machine learning as a kernel method}
Conventional variational quantum machine learning models \cite{mitarai2018quantum,Schuld2019feature,havlivcek2019supervised} work in the following manner.
First, data are encoded into quantum states by $U_{\phi}(\bm{x})$.
Then, we apply a trainable parameterized circuit $V(\bm{\theta})$.
Finally, we measure the expectation value of an observable $O$, which is used as the model output $y(\bm{x},\bm{\theta})$. 
Mathematically, the above process can be written as,
\begin{align}
    y(\bm{x},\bm{\theta}) = \langle 0^n |U_{\phi}^{\dagger}(\bm{x}) V^{\dagger}(\bm{\theta}) O V(\bm{\theta}) U_{\phi}(\bm{x}) | 0^n \rangle \, .
\end{align}
The correspondence of this model to a linear model can be readily seen by rewriting the above expression as,
\begin{align}\label{eq:density-matrix-feature-linear-model}
    y(\bm{x},\bm{\theta}) = \mathrm{Tr}(O(\bm{\theta})\rho(\bm{x})),
\end{align}
where
\begin{align}
    O(\bm{\theta}) &= V^{\dagger}(\bm{\theta}) O V(\bm{\theta}) \, ,\\
    \rho(\bm{x}) &= U_{\phi}(\bm{x}) | 0^n \rangle\bra{0^n}U_{\phi}^\dagger(\bm{x})
\end{align}
Since $\mathrm{Tr}(A^\dagger B)$ for operators $A$ and $B$ defines an inner product in the operator space, Eq. \eqref{eq:density-matrix-feature-linear-model} defines a linear model using the feature vector $\rho(\bm{x})$ and the weight vector $O(\bm{\theta})$ \cite{schuld2021supervised}.
Therefore, if we construct a kernel-based model using the same feature vector $\rho(\bm{x})$, it is guaranteed to achieve better performance on a training dataset than the model in Eq. \eqref{eq:density-matrix-feature-linear-model}.
In this case, we define the kernel function as,
\begin{align}\label{conventional-kernel}
    K_{q}(\bm{x}_i, \bm{x}_j) &= \mathrm{Tr}(\rho(\bm{x}_i)\rho(\bm{x}_j))\\
    &= | \langle \phi(\bm{x}_i) | \phi(\bm{x}_j) \rangle |^2 \, ,
\end{align}
where $| \phi(\bm{x}) \rangle$ is defined as follows
\begin{align}
    | \phi(\bm{x}) \rangle = U_{\phi}(\bm{x}) |0^n \rangle .
\end{align}
We call the kernel methods that are based on $K_q(\bm{x}_i,\bm{x}_j)$ the conventional quantum kernel method.
% ここらへんの話あんまり関係ない気がしたのでコメントアウト
%According to the Representer theorem, an optimal model is given by

%\begin{align}
    %f_{\rm opt}(\bm{x}) = \sum^{N}_{i=1} \beta_i K(\bm{x}_i, \bm{x}) \, .
%\end{align}
%where $\bm{x}_i$ is the training data.
%This means that finding optimal measurement corresponds to optimizing coefficients $\beta_i$.
%If the loss function is convex, we can find the globally optimal measurement in the $N$ dimensional subspace.
%However, in the case of variational quantum circuit, its measurement is parameterized with $\bm{\theta}$.
%$\bm{\theta}$-subspace does not always contain the globally optimal measurement.
%
The above argument only holds for a quantum model $y(\bm{x},\bm{\theta}) = \bra{0}U^\dagger(\bm{x},\bm{\theta}) O U(\bm{x},\bm{\theta})\ket{0}$ with $U(\bm{x},\bm{\theta})$ in the form of $V(\bm{\theta}) U_{\phi}(\bm{x})$.
We propose quantum circuits that cannot be splitted in such a way in Sec. \ref{sec:numerical_experiment}.
This modification prevents us from rewriting the model into the form of Eq. \eqref{eq:density-matrix-feature-linear-model}, and thus we cannot construct an equivalent kernel model.

\section{Quantum Tangent Kernel}

In this section, we apply the formulation of the NTK described in Sec.\ref{subsec:NTK} to parameterized quantum circuits.
%discuss the training dynamics of an over-parameterized quantum cicuit and propose Quantum Tangent kernel (QTK).
%The width of a quantum circuit corresponds to the number of qubits, but it is practically difficult to take the limit of a large width. 
%For this reason, we consider an over-parameterized quantum circuit with a large number of layers. 
We consider a model whose output is given as the expectation value of an operator $O$ as,
\begin{align}
    y(\bm{x},{\bm{\theta}}) = \langle 0^n | U^{\dagger}(\bm{x}, \bm{\theta}) O U(\bm{x}, \bm{\theta}) | 0^n \rangle \, ,
\end{align}
where $\bm{x}$ is the input data and $\bm{\theta}$ are parameters of a quantum circuit.
We define the following kernel by using the output of a quantum circuit analogous to the NTK.
\begin{align}\label{eq:QTK}
    K_{\rm qtk}(\bm{x}_i, \bm{x}_j) = \nabla_{\theta} y(\bm{x}_i,\bm{\theta}_0)^T \nabla_{\theta} y(\bm{x}_j,\bm{\theta}_0) \, .
\end{align}
We call the kernel $K_{\rm qtk}(\bm{x}_i, \bm{x}_j)$ quantum tangent kernel (QTK).
As in NTK, QTK is valid as long as the parameters do not change much from their initial random guess $\bm{\theta}_0$ when, for example, updating them by gradient descent based with respect to some suitable cost function.

\begin{figure}
    \centering
    \includegraphics[scale=0.27]{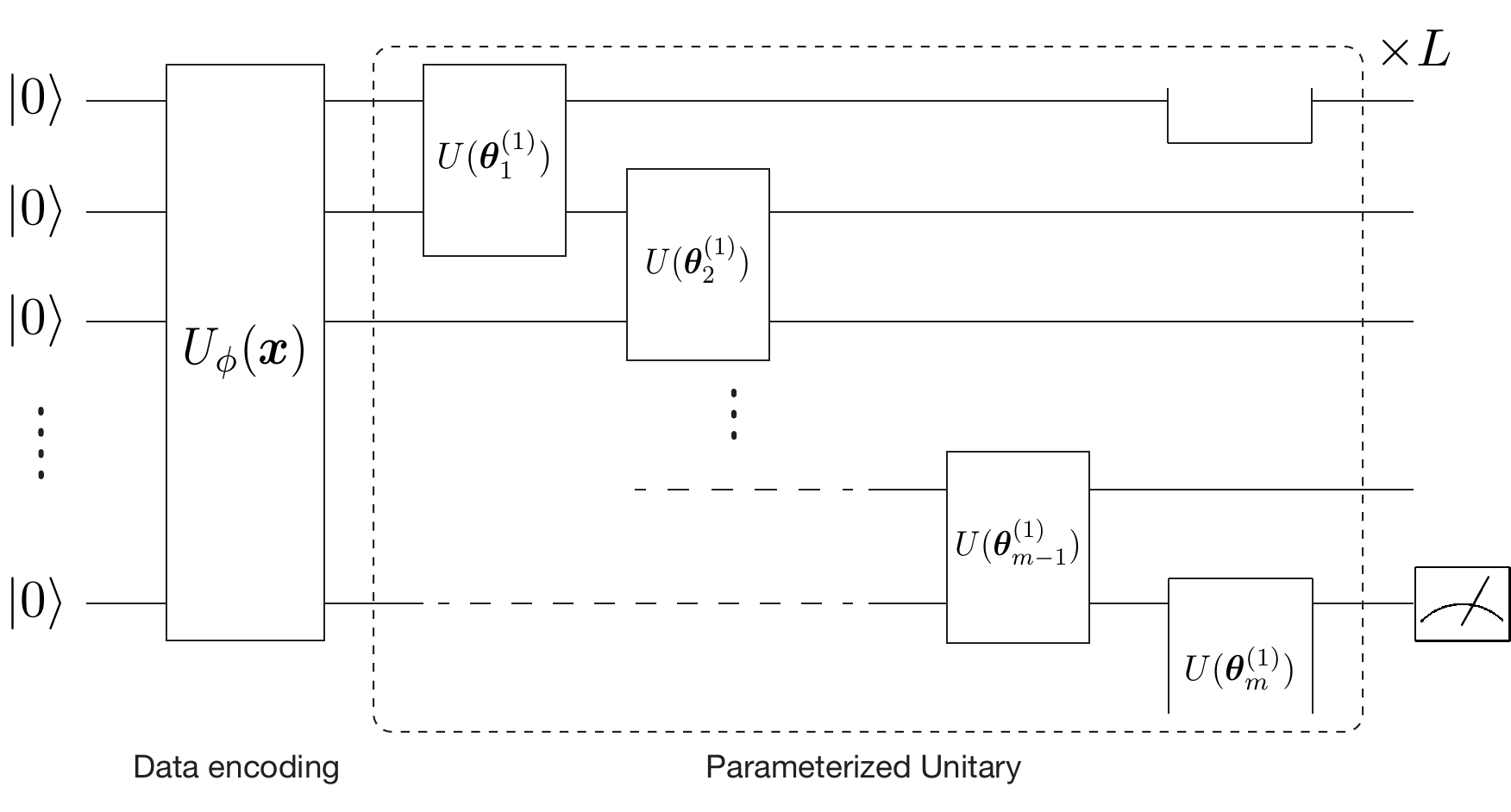}
    \caption{The $m$-qubit ansatz used for numerical simulations.
    $U_{\phi}(\bm{x})$ is the quantum feature map for encoding classical data. $U(\bm{\theta}_j^{(i)}) \in SU(4)$ is the parameterized unitary of the $i$-th layer.
    At the output, we measure the Pauli $Z$ expectation value of the final qubit.
    }
    \label{fig:qntk-circuit}
\end{figure}

\begin{figure}
    \centering
    \includegraphics[scale=0.33]{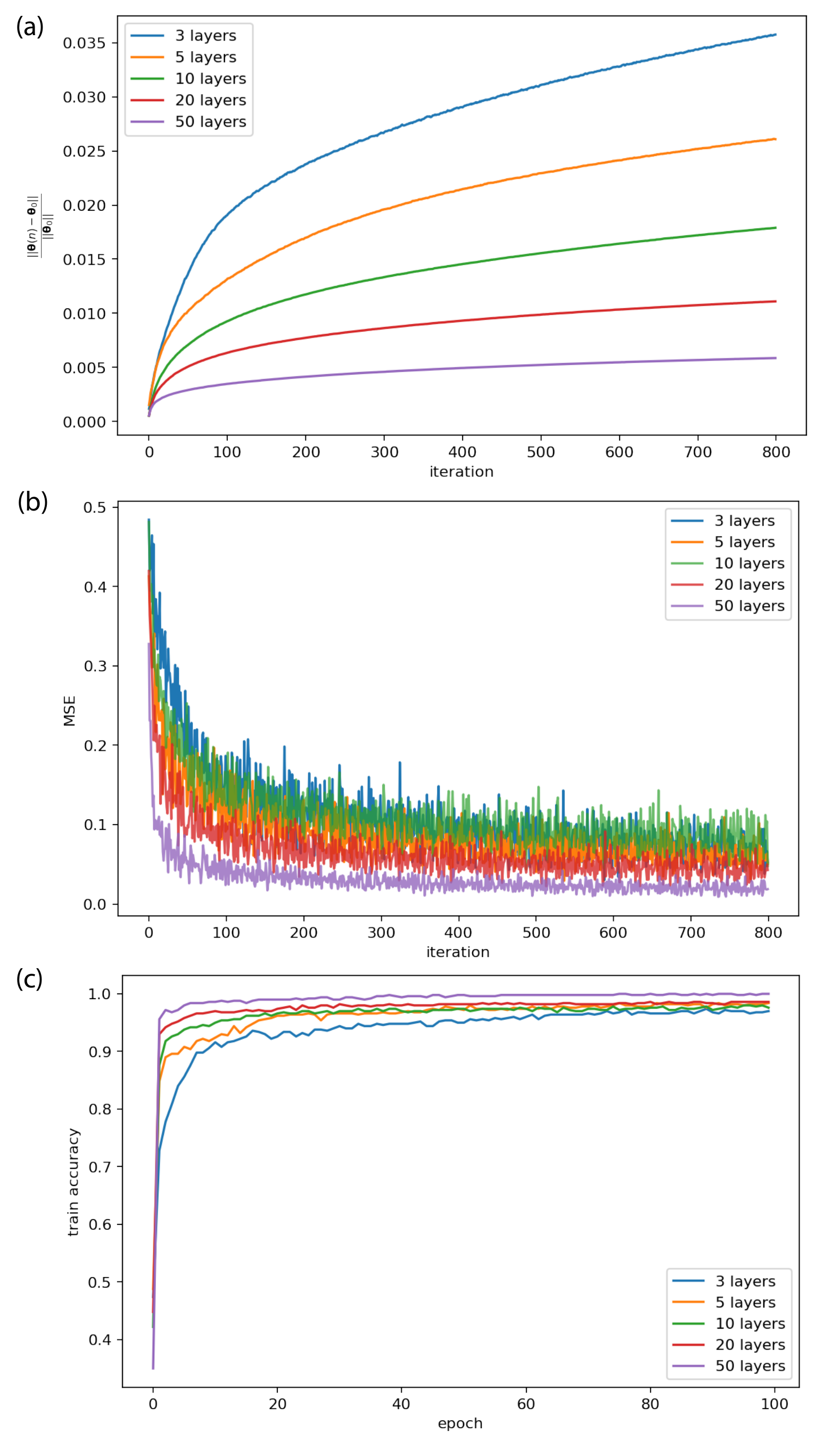}
    \caption{(a) Relative norm change in the parameters of the quantum circuit from initial values during training by gradient descent. $\bm{\theta}(n)$ is the parameter at the $n$-th iteration. $\bm{\theta}_0$ is initial value of the parameter. (b)(c) Behavior of training losses and training accuracies for different number of layers during training.}
    \label{fig:qntk-parameter-relative-change}
\end{figure}

We can numerically check that such a phenomena occurs for quantum circuits with a large number of layers. 
To this end, we train quantum circuits with varying numbers of layers and look at the changes of their parameters.
We train a 10-qubit circuit as shown in Fig.~\ref{fig:qntk-circuit} on the MNIST dataset \cite{deng2012mnist} reduced to a 10-dimensional datasets by principal component analysis (PCA) and use only the digits 8 and 3 for binary classification.
In this numerical experiment, the quantum feature map to encode the data is given by $U_{\phi}(\bm{x}) = \bigotimes_{i = 1}^{10} \exp(i x_i Y_i)$.
The initial values of the parameters of these quantum circuits are set randomly using a uniform distribution between $0$ and $2 \pi$.
We use the mean squared error (MSE) as the loss function and the standard stochastic gradient descent with mini-batch size 64 to train the quantum circuits.
The expectation value of the deep quantum circuit concentrate on its average over Hilbert space due to barren plateau issue \cite{mcclean2018barren}.
In our case, the expectation value of the observable $O = Z_{10}$ on the deep quantum circuit concentrate on 0.
Now we use MSE as the loss function and the target value is 1 or -1 in the classification problem.
Even if the classification is correct, if the expectation value is close to 0, the MSE will not decrease.
For this reason, we multiply a scale factor to the expectation value, i.e., 
$y(\bm{x},{\bm{\theta}}) = C \langle 0^n | U^{\dagger}(\bm{x}, \bm{\theta}) Z_{10} U(\bm{x}, \bm{\theta}) | 0^n \rangle $. 
The variance of the expectation value of the quantum circuit multiplied by a scale factor should be roughly 1.0 to decrease the MSE.
Fig.~4 in \cite{mcclean2018barren} is the case of the gradient, it is also an expectation value of a deep quantum circuit.
In this experiment, the number of layers in a quantum circuit is at most 50.
Thus, we can roughly estimate the value of variance of the expectation value before multiplying a scale factor to be between $O(10^{-1})$ and $O(1)$.
We choose $C=4.0$ to decrease the MSE and verify that the training is successful.
Fig.~\ref{fig:qntk-parameter-relative-change} (a) shows the relative norm changes in the parameters of the quantum circuits with layers $L = 3, 5, 10, 20$ and $50$.
We can observe that when a quantum circuit has more layers, the changes of the parameters during training becomes small.
Figs.~\ref{fig:qntk-parameter-relative-change} (b) and (c), which respectively show the decrease of MSE and the increase of training accuracy, indicate the success of training.
This result implies that it is possible to linearly approximate $y(\bm{x},\bm{\theta})$ with respect to its parameters when the circuit is sufficiently deep.
Hence, we can expect QTK to provide a machine learning model that is approximately equivalent to $y(\bm{x},\bm{\theta})$.

For a given ansatz, the QTK can be calculated on a quantum computer with parameter-shift rules \cite{mitarai2018quantum, schuld2019evaluating} and its generalizations \cite{Banchi2021measuringanalytic,wierichs2021general}.
This provides us an alternative quantum machine learning model other than the conventional variational methods and quantum kernel methods.
Note that, as opposed to the case of the NTK \cite{DBLP:journals/corr/abs-1806-07572,DBLP:journals/corr/abs-1904-11955}, currently we do not know how to calculate the QTK analytically for a given form of ansatz in the infinite size limit.
We leave this direction as an interesting future direction to explore.

% \begin{align}
%     \frac{\partial f_{\rm qtk}(\vector{x}, \vector{\theta})}{\partial \theta_i} = 
%     f_{\rm qtk}  \left (\vector{x}, \vector{\theta} + \frac{\pi}{4} \vector{\Delta}_i \right ) - f_{\rm qtk} \left (\vector{x}, \vector{\theta} - \frac{\pi}{4} \vector{\Delta}_i \right ) \, ,
% \end{align}
% where $\vector{\Delta}_i$ is a unit vector. its $i$-th component is one and all others are zero.

\section{Numerical Experiment}\label{sec:numerical_experiment}
%\subsection{Linear approximation of the output of over-parameterized quantum circuit}

\begin{figure}
    \centering
    \includegraphics[scale=0.27]{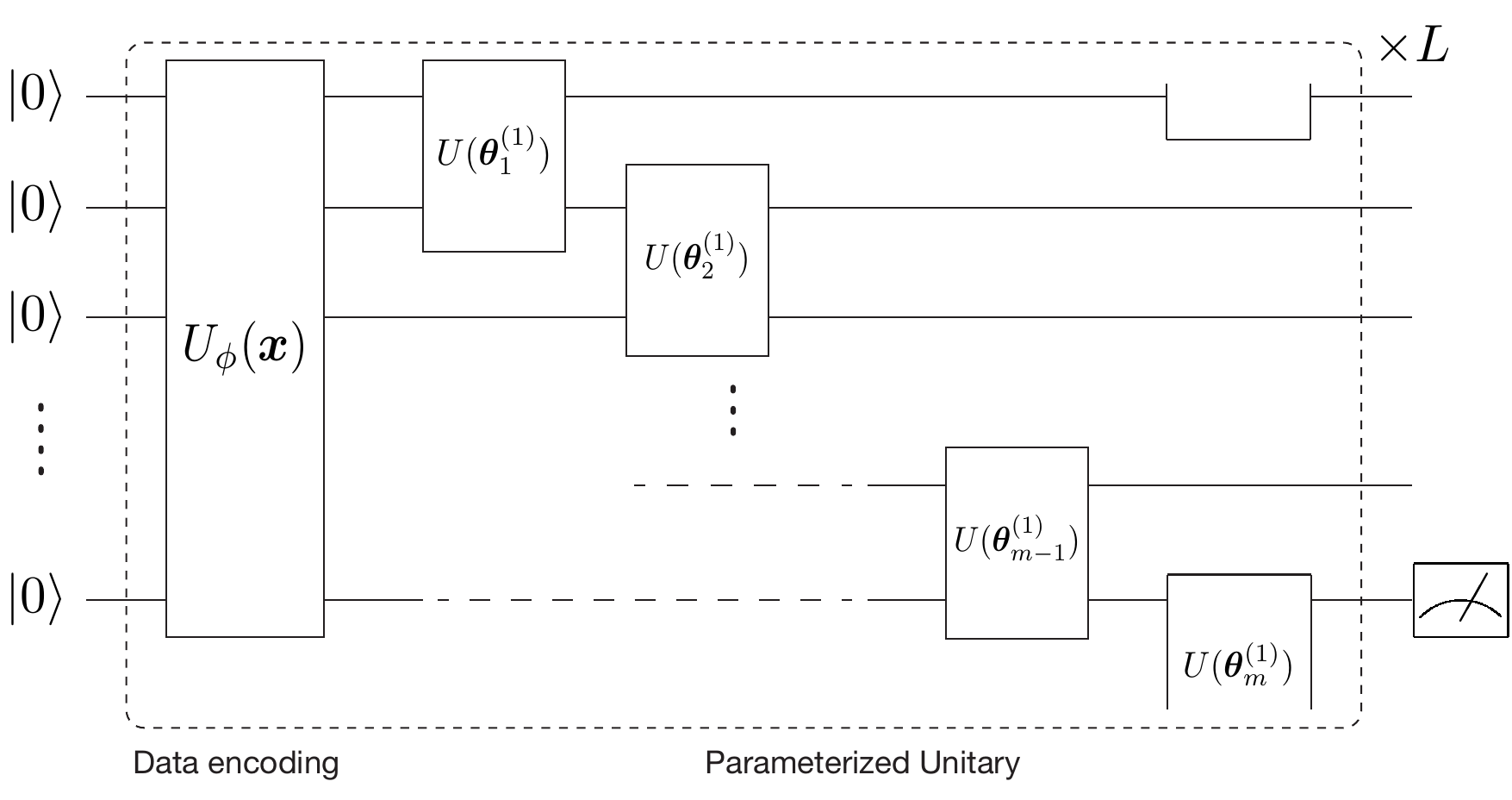}
    \caption{
    Quantum circuit that defines deep quantum tangent kernel. $U_{\phi}(\bm{x})$ is the feature map and $U(\bm{\theta}_j^{(i)}) \in SU(4)$ is the parameterized unitary of the $i$-th layer. To increase the non-linearity of the kernel, the feature map $U_{\phi}(\bm{x})$ and the parameterized unitary are iteratively applied.
    At the output, we measure the Pauli $Z$ expectation value of the final qubit.
    }
    \label{fig:deep-qntk-circuit}
\end{figure}

We numerically demonstrate the expressive power of quantum tangent kernel.
Here, we consider two types of quantum tangent kernel according to how the data is encoded into quantum states.
First one is a circuit where $\bm{x}$ is encoded only at the first layer as in Fig.~\ref{fig:qntk-circuit}:
\begin{align}\label{eq:shallow QTK ansatz}
    U_{\mathrm{shallow}}(\bm{x}, \bm{\theta}) =  V(\bm{\theta}) U_{\phi}(\bm{x}) \, ,
\end{align}
where $V(\bm{\theta})$ is a parameterized unitary 
%with a large number of layers to over-parameterize a quantum circuit
and $U_{\phi}(\bm{x})$ is a quantum feature map to encode data. 
This type of quantum circuit can be interpreted as the conventional quantum kernel method.
We call the QTK associated with this type of ansatz shallow QTK.
%We now consider the $m$-qubit parameterized quantum circuit as shown in Fig.\ref{fig:qntk-circuit} and its parameterized unitary is given by
%\begin{align}
%    U_{\rm qtk}(\vector{\theta}) = \prod_{i=1}^{L} \prod_{j=1}^{m} U(\vector{\theta}_j^{(i)}) \, .
%\end{align}
%where $U (\vector{\theta}_i^{(j)} ) \in SU(4)$ is two-qubit unitary gate and $L$ is the number of layers.
%As in the case of NTK, the parameters in the over-parameterized quantum circuit are supposed to change only slightly from their initial values and linear approximation of $f(\vector{x}, \vector{\theta})$ is valid.
% Next, we consider the second type of quantum tangent kernel which cannot be interpreted as the kernel method in the sense described above.
%In machine learning, the kernel method is a mapping to a higher dimensional feature space to extract the non-linearity of features.
% Kernel methods maps a datum to a higher dimensional feature space to construct models that are nonlinear with respect to the original data.
%However, the data encoding described above can only introduce non-linearity from the tensor product structure between qubits.
Next, in order to increase the non-linearity, we consider a multi-layered circuit that alternates between data encoding and a parameterized unitary as shown in Fig.~\ref{fig:deep-qntk-circuit}.
More concretely, we consider the following unitary: 
%\begin{align}
%    f_{\rm dqtk}(\vector{x}, \vector{\theta}) = \langle 0^n | U_{\rm dqtk}(\vector{x}, \vector{\theta}) Z_m U_{\rm %dqtk}^{\dagger}(\vector{x},  \vector{\theta}) | 0^n \rangle
%\end{align}
%\begin{align}
%    U_{\rm dntk} (\vector{x}, \vector{\theta}) = \prod_{i=1}^{n} U_{\phi}(\vector{x}) %U(\vector{\theta}_i) \, .
%\end{align}
\begin{align}\label{eq:deep QTK ansatz}
    U_{\mathrm{deep}}(\bm{x}, \bm{\theta}) = \prod_{i=1}^{L} \left [ V(\bm{\theta}_i) U_{\phi}(\bm{x}) \right ] \, .
\end{align}
%where $U_j(\vector{\theta}_i^{(j)}) \in SU(4)$  is two-qubit unitary gate and $U_{\phi}(\vector{x})$ is a quantum feature map to encode data.
We call the QTK associated with this type of ansatz deep QTK. The same type of ansatz is proposed in Ref. \cite{jerbi2021quantum}.

Since $U_{\mathrm{deep}}$ has higher non-linearity, it is expected to show higher expressive power than shallow QTK.
In order to demonstrate the expressive power, we generate a ``ansatz-generated'' dataset by using a quantum circuit for $U_{\mathrm{deep}}$ and classify the data using SVM with the three types of kernels: shallow QTK, deep QTK and conventional quantum kernel defined by the feature map $U_{\phi}(\bm{x})$.
If this data is classified efficiently by SVM with deep QTK, it has higher expressive power than the other two.

In this numerical experiment, we use $U_{\bm{\phi}}(\bm{x})$ in the form of,
\begin{align}
    U_{\phi}(\bm{x}) = \exp  \left (i \sum_{S \subseteq [m] } \phi_S(\bm{x}) \bigotimes_{i \in S} Y_i \right ) \, ,
\end{align}
where $Y_i$ is the Pauli $Y$ operator acting on the $i$-th qubit.
% $\vector{x}$ is classical data and $\phi_S(\vector{x}) \in (0, 2 \pi ] $ is encoded as a coefficient of Pauli $Y$.
In our experiment, interaction up to 2-qubit $|S| \leq 2$ is given and acts on the initial state $|0^n \rangle$.
This type of quantum feature map has been proposed in Ref.~\cite{havlivcek2019supervised}.
The functions $\phi_i(\bm{x})$ and $\phi_{ij}(\bm{x})$ are given by,
\begin{align}
    \phi_{i}(\bm{x}) &= \arcsin (x_i) \, , \\
    \phi_{ij}(\bm{x}) &= \arcsin{(x_i x_j)} \, .
\end{align}
%質問
The ``ansatz-generated'' dataset is generated in the following manner.
Four dimensional random value data $\{\bm{x}_i\}$ are inputted into $U_{\mathrm{deep}}(\bm{x},\bm{\theta})$ consisting of $n=4$ qubits and $L = 10$ layers.
Then, we evaluate the expectation value
\begin{align}\label{eq:dee-qtk-circuit}
    l(\bm{x}_i, \bm{\theta}) = \langle 0^n | U^{\dagger}_{\rm deep}(\bm{x}_i, \bm{\theta}) Z_4 U_{\rm deep}(\bm{x}_i, \bm{\theta}) | 0^n \rangle \, %,\\
    % U_{\rm deep} (\vector{x}, \vector{\theta}) = \prod_{i}^L \left [ \prod_{j}^m U(\vector{\theta}_j^{(i)}) U_{\phi}(\vector{x}) \right ] \, .
\end{align}
with a randomly chosen $\bm{\theta}$.
% We prepare the target data for the classification task 
We label each $\bm{x}_i$ as 1 and -1 if $l(\bm{x},\bm{\theta})\geq 0$ and $l(\bm{x},\bm{\theta})< 0$, respectively.
%We generate 15,000 data and split them into 1000 training and 5000 test data.
We generate the 15,000 samples of the four dimensional input data $\bm{x}_i$ and its label $y_i$.
They are splitted into 10,000 training and 5000 test data.

%$\{ (\vector{x}_i, y_i) \}$ ansatz-generated data set are generated with its outputs as labels.
%We generate ansatz-generated data set and split them into training and test data. 
%input random value data into a quantum circuit for deep QTK and use its outputs as labels

We classify the above dataset with SVM using three different kernels: shallow QTK [Eq.~\eqref{eq:QTK} with $U(\bm{x},\bm{\theta})=U_{\mathrm{shallow}}(\bm{x},\bm{\theta})$ and $L=10$ layers], deep QTK [Eq.~\eqref{eq:QTK} with $U(\bm{x},\bm{\theta})=U_{\mathrm{deep}}(\bm{x},\bm{\theta})$ and $L=10$ layers], and conventional quantum kernel [Eq.~\eqref{conventional-kernel}]. %$K(\bm{x}_i,\bm{x}_j)=\bra{0}U^\dagger_\phi (\bm{x}_i)U^\dagger_\phi(\bm{x}_j)\ket{0}$.
In order to calculate QTK and deep QTK, we randomly set parameter values of a quantum circuit using a uniform distribution between $0$ and $2 \pi$.
Note that the parameters used in this learning phase is different from the one used for data generation.
We apply $U(\bm{\theta}_i^{(j)}) \in SU(4)$ on each neighboring qubits. $SU(4)$ is parameterized according to Cartan decomposition as follows
\begin{align}
& U(\bm{\theta}_i^{(j)}) = \\ \nonumber
    & k_1 \exp  \left [ \frac{i}{2} (\theta_{xx} X_i X_{i+1} + \theta_{yy} Y_i Y_{i+1} + \theta_{zz} Z_i Z_{i+1} ) \right ] k_2 \, .
\end{align}
where $k_1, k_2 \in SU(2) \otimes SU(2)$ and $X_i$, $Y_i$, $Z_i$ are the Pauli operators.
QTKs are calculated by the parameter-shift rule \cite{mitarai2018quantum,schuld2019evaluating}.
The regularization strength is determined via cross-validation for each kernel.

The results of the classification task are listed in Table~\ref{table:accuracy-classification}.
Among three kernels, deep QTK outperforms other kernels.
This result indicates that deep QTK employs a feature map that cannot be expressed very accurately with other kernels.
In contrast, shallow QTK is essentially just a quantum kernel method using the feature map in Eq.~\eqref{eq:shallow QTK ansatz}, so its performance is not improved compared to conventional quantum kernel as expected.

\begin{figure}[htbp]
    \centering
    \includegraphics[scale=0.42]{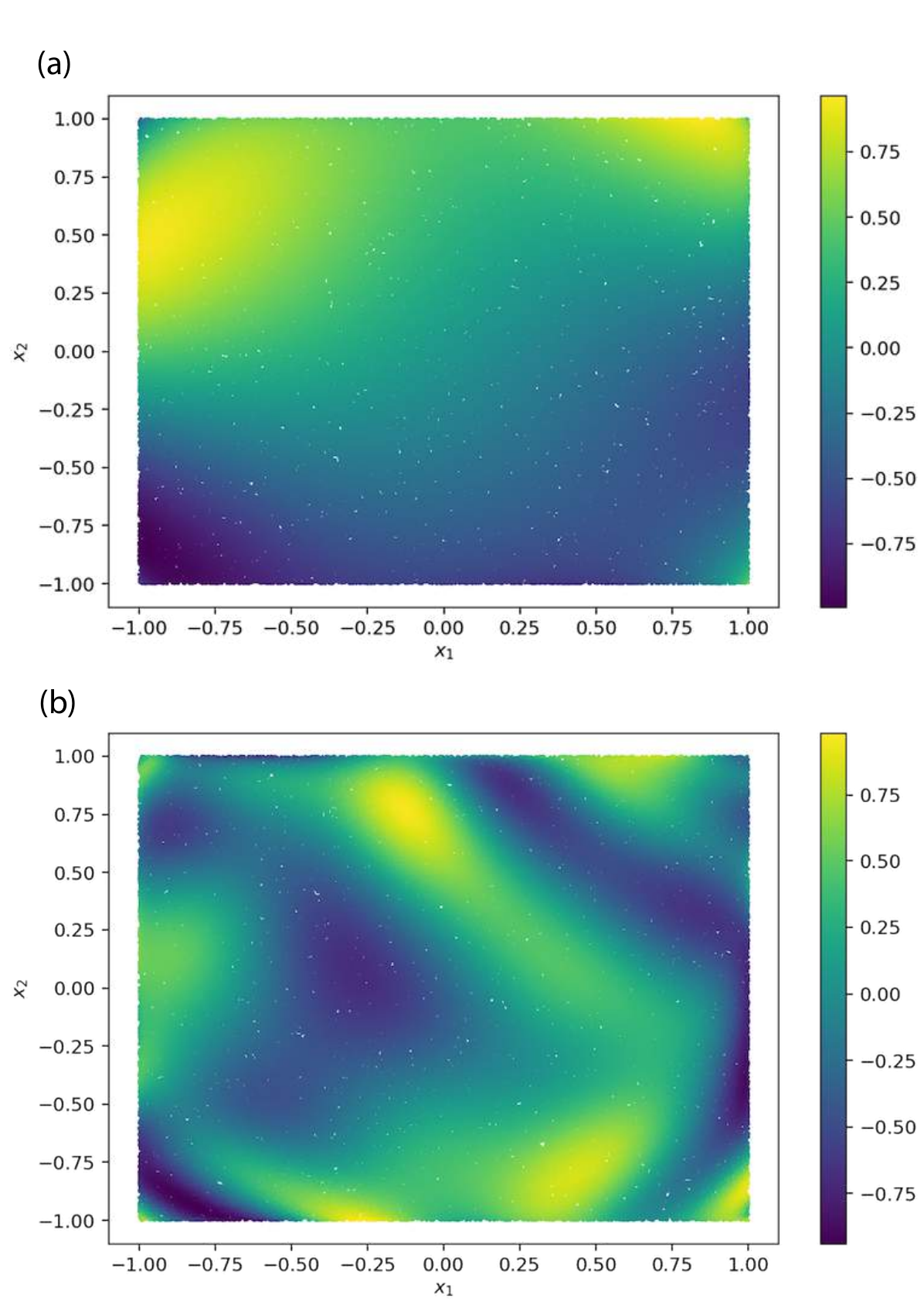}
    \caption{(a) The distribution of outputs of the quantum circuit with $L = 10$ layers (Fig.~\ref{fig:qntk-circuit}) which can be interpreted as the conventional quantum kernel method. (b) The distribution of outputs of the quantum circuit with $L = 10$ layers (Fig.~\ref{fig:deep-qntk-circuit}) beyond the conventional quantum kernel.}
    \label{fig:qtk-random-output}
\end{figure}

The expressive power of kernels can be illustrated more directly by visualizing the feature map of each kernel.
Fig.~\ref{fig:qtk-random-output} shows the distribution of the ansatz-generated dataset generated by quantum circuits for $U_{\mathrm{shallow}}(\bm{x},\bm{\theta})$ and $U_{\mathrm{deep}}(\bm{x},\bm{\theta})$.
In order to visualize the distribution, we generate two dimensional data using the two-qubit quantum circuits. 
%We use the following feature maps $\phi_i(\vector{x})$ and $\phi_{ij}(\vector{x})$ when the data are encoded.
%Next, we demonstrate the performance of the quantum tangent kernel for classification tasks using SVM and expression power of deep quantum. 
%\begin{align}
%    \phi_{i}(\vector{x}) &= \arcsin (x_i) \, , \\
%    \phi_{ij}(\vector{x}) &= \arcsin{(x_i x_j)} \, .
%\end{align}
% By using $\arcsin$, the oscillations of outputs caused by the quantum feature map is canceled out and it becomes easier to check the non-linearity caused by the structure of the quantum circuit.
% In Fig. \ref{fig:qtk-random-output}, (a) and (b) shows the distribution of outputs of a quantum circuit for QTK and deep QTK.
These distributions show that $U_{\mathrm{deep}}(\bm{x},\bm{\theta})$ expresses higher non-linearity than $U_{\mathrm{shallow}}(\bm{x},\bm{\theta})$ does.

\begin{center}
\begin{table}
\centering
\begin{tabular}{cc} \hline
Kernel & Accuracy \\ \hline
Quantum kernel  & 0.7842 \\
Shallow quantum tangent kernel & 0.7484 \\
Deep quantum tangent kernel & 0.812 \\ \hline
\end{tabular}
\caption{Classification accuracy for SVM with three types of kernels. Three SVMs classify the ansatz-generated dataset generated by a quantum circuit for deep QTK as shown in Fig.\ref{fig:deep-qntk-circuit}.}
\label{table:accuracy-classification}
\end{table}
\end{center}

\section{Conclusion}
We proposed quantum tangent kernel (QTK) and deep quantum tangent kernel which cannot be interpreted as conventional quantum kernel methods described in Ref.~\cite{schuld2021supervised}.
QTK is defined by applying the formulation of NTK to parametrized quantum circuits.
We find that parameters of an overparameterized quantum circuit change only slightly from its initial values during training. 
This indicates that the output of an overparameterized quantum circuit can be linearly approximated, which validates the formulation of QTK.
By using this overparameterization, we can avoid the gradient descent for quantum circuits with a large number of parameters and easily optimize the parameters.
Then, in order to increase the non-linearity of a feature map, we introduced a multi-layered data encoding that alternates between data encoding and a parameterized unitary. 
This encoding method increases non-linearity of the feature map and improves the expressive power of kernels.
% The kernel using this type of the quantum circuit is called deep QTK.

We demonstrated the performance of shallow QTK and deep QTK for the classification task. Using the ansatz-generated dataset generated by the quantum circuit for deep QTK (Eq.~\eqref{eq:deep QTK ansatz}), we evaluate the performance by using SVM with three kernels: shallow QTK, deep QTK and conventional quantum kernel. 
We showed that SVM with deep QTK outperforms SVM with other kernels and deep QTK has a feature map with high non-linearity.

Our results imply that deep parameterized quantum circuits with repetitive data encoding unitary have a higher representation power and better performance for quantum machine learning
than the conventional quantum kernel method.
While we here employed an overparameterization limit and hence a deeper quantum circuit to take the neural tangent kernel approach,
sophisticatedly trained neural networks, such as deep neural networks, provide a better performance in general than neural tangent kernels
as known in the classical literatures \cite{li2019enhanced}.
It gives us hope that the real fruit of quantum machine learning is not in the shallow or deep limit, but in the mild depth, which could be modeled by neither conventional nor tangent kernel methods.
Therefore, better ansatz constructions and parameter optimization methods are crucially important.

% An important future direction to explore is...

\section*{Acknowledgments}
KM is supported by JST PRESTO Grant No. JPMJPR2019 and JSPS KAKENHI Grant No. 20K22330.
KF is supported by JST ERATO JPMJER1601, and JST CREST JPMJCR1673.
This work is supported by MEXT Quantum Leap Flagship Program (MEXT QLEAP) Grant Number JPMXS0118067394 and JPMXS0120319794.
We also acknowledge support from JST COI-NEXT program.

\bibliographystyle{apsrev4-1}
\bibliography{references}

\end{document}